\documentclass[usegraphicx]{mn2e}
\usepackage{times}

\newif\ifAMStwofonts

\newcommand{\gtorder}{\mathrel{\raise.3ex\hbox{$>$}\mkern-14mu
                \lower0.6ex\hbox{$\sim$}}}
\newcommand{\ltorder}{\mathrel{\raise.3ex\hbox{$<$}\mkern-14mu
                \lower0.6ex\hbox{$\sim$}}}

\newcommand{\ovi}{O\,{\sc vi}}
  
\newcommand{\heii}{He\,{\sc ii}}

\newcommand{\hi}{H\,{\sc i}}  
 
\newcommand{\civ}{C\,{\sc iv}} 
 
\newcommand{\nv}{N\,{\sc v}}

\newcommand{\mgii}{Mg\,{\sc ii}} 
 
\newcommand{\apj}{ApJ}
\newcommand{\aj}{AJ}
\newcommand{\apjs}{ApJS}

\newcommand{\mnras}{MNRAS}

\title{Non-Sobolev modelling of radiation pressure driven flows in Active Galactic Nuclei.}
\author[Doron Chelouche and Hagai Netzer]
       {Doron Chelouche\thanks{email: doron@wise.tau.ac.il; netzer@wise.tau.ac.il} and Hagai
         Netzer\mbox{\raise.9ex\hbox{$\star$}} \\
        School of Physics and Astronomy and the Wise Observatory,
        The Beverly and Raymond Sackler Faculty of Exact Sciences,\\
        Tel Aviv University, Tel Aviv 69978, Israel}

\pubyear{2000}

\begin{document}
\maketitle

\label{firstpage}

\begin{abstract}
We present a new general scheme for  calculating the structure and dynamics of radiation pressure driven, photoionized flows. The new method goes one step beyond the Sobolev approximation. It involves a numerical solution of the radiative transfer in absorption lines including the effects of differential expansion and line interactions such as line locking and blanketing. We also present a new scheme for calculating  the radiation pressure due to trapped line photons in finite, differentially expanding flows. We compare our results for the radiation pressure force with those obtained using the Sobolev approximation and show the limitations of the latter. In particular, we  demonstrate that the Sobolev method gives a poor approximation near discontinuity surfaces and its neglect of line blanketing can lead to erroneous results in high velocity flows. We combine the newly calculated radiation pressure force with self-consistent photoionization and thermal calculations to study the dynamics and spectral features of broad absorption line flows and highly ionized gas flows in AGN. A comparison with Sobolev-type calculations shows that the latter over estimates the flow's terminal velocity and, conversely, under estimates its opacity. We also show that line locking on broad emission lines can have a significant effect on the dynamics and spectral features of AGN flows. 
\end{abstract}


\section{Introduction}

Spectroscopic observations of type-I (broad line) active galactic nuclei (AGN) show a myriad of UV and X-ray absorption lines  which, in many objects, are blueshifted relative to the emission lines, implying outflows with a wide range of  velocities. Such outflows can be crudely divided into narrow absorption line (NAL) flows and broad absorption line (BAL) flows. NAL flows are common in Seyfert 1 galaxies and are characterized by low velocities ($\ltorder 1000~{\rmn{km~s}^{-1}}$) and narrow absorption profiles (full width at half maximum, FWHM$\sim 100~{\rmn{km~s}^{-1}}$). BAL flows are observed in $\sim 10\%$ of luminous AGN (BALQSOs, which also show NAL features) showing sub-relativistic velocities ($\ltorder 20,000~{\rmn{km~s}^{-1}}$) and FWHM of the same order. The location and mass of the gas responsible to the observed NAL and BAL flows are poorly constrained by observations (e.g., Kraemer et al. 2001, Netzer et al. 2002, de Kool et al. 2002 and Everett, Koenigl, \& Arav 2002). Theoretical modeling is therefore important for understanding the origin and the mass flux associated with such flows.

The modelling of AGN outflows stems from the early studies of stellar flows involving acceleration by radiation pressure force (Castor Abbott, \& Klein 1975, hereafter CAK; but see e.g., Blandford 1994 and for additional types of models). Stellar winds are assumed to be in LTE and are driven by radiation pressure in some $10^5$ lines.  Most works in this area used the Sobolev approximation  to calculate the radiation pressure force in the lines (CAK). The method has also been used to model AGN flows (e.g., Arav, Li, \& Begelman 1994, Murray et al. 1995, and Proga, Stone, \& Kallman 2001) despite the very different physical conditions. However, there are several complications associated with AGN flows which render Sobolev like methods unwarranted. For example, line overlap (line blanketing and locking; e.g., Arav 1996) in high velocity flows cannot be modeled by Sobolev type methods. A different complication arises in NAL flows where the line width is a few times the thermal width (e.g., Crenshaw et al. 1999), and cannot be treated by the Sobolev approximation. Thus, self-consistent solution of AGN flows requires  more accurate transfer methods. 

There are several other problems unique to AGN flows that must be treated in a more self-consistent way. One example is the internal radiation pressure resulting from multiple scattering of resonance line photons. The process has been described and analyzed by Elitzur \& Ferland (1986; hereafter EF86), yet its application to realistic flows requires some modification.

This work is the first in a series of papers that  aim  to correctly model the structure of radiation pressure driven, photoionized flows in AGN.  In this paper we present a new scheme we have developed and elaborate on the parts that are significantly different from previous results. The new method takes into account line overlap and uses a better treatment of the radiation pressure force close to discontinuity surfaces. We also revise and improve the EF86 scheme for the calculation of internal radiation pressure. The new  scheme results in  different ionization structure, flow dynamics, and different absorption line profiles. The following papers will apply the scheme to several important physical situations, such as shielded flows, and will investigate specific issues pertaining to the dynamics, geometry and thermal stability of AGN flows.

The paper is organized as follows: In section 2 we present the new scheme for calculating AGN flows. In section 3 we demonstrate the improved capabilities  by comparing the results to those obtained by the Sobolev approximation and study the dynamics and spectral features in NAL and BAL AGN flows. We summarize our results in section 4.

\section{Method}

In this section we present the new numerical scheme and emphasize the improved results for the radiation pressure, radiation pressure force, and the absorption line profiles.

\subsection{The equation of motion}

The equation of motion for radiatively driven flows is
\begin{equation}
v\frac{dv}{dr}=\frac{1}{\rho} \left ( F_{\rmn{rad}} - \frac{dP_{\rmn{tot}}}{dr} \right )-g(r),
\label{eqnmot}
\end{equation}
where $v$ is the flow velocity, $r$ the radial coordinate, $\rho$ the flow density, $F_{\rmn{rad}}$ the radiation pressure force, $P_{\rmn{tot}}$ the total (gas and radiation) pressure, and $g$ the gravity. We assume that contributions due to other forces, e.g., drag force, are negligible. This equation is solved in conjunction with a continuity condition $\rho=\rho (r,v)$ which is defined for the geometry under consideration. The important issues of critical points and flow continuity are addressed at the end of this section and in appendix B.

\subsection{The radiation pressure force}

We are interested in flows whose dynamics is dominated by radiation pressure force. To express this force, we make use of the force multiplier formalism (see CAK, Stevens \& Kallman 1991, and Chelouche \& Netzer 2001; hereafter CN01) which assumes that
\begin{equation}
F_{\rmn rad}=H_c M,
\label{mdef}
\end{equation}
where
\begin{equation}
H_c=\frac{n_e \sigma_T L_{\rmn{tot}}}{4\pi r^2c}
\label{hcdef}
\end{equation}
is the Compton factor, $L_{\rmn{tot}}$ is the bolometric luminosity of the source, $\sigma_T$ the Thomson cross section, $n_e$ the free electron density, and $r$ the radial coordinate. We note that for relatively neutral gas, the definition of $H_c$ is different and must include Rayleigh scattering of low energy photons and Compton scattering of X-ray photons on bound electrons. Here we focus on ionized gas for which equation \ref{hcdef} holds. We also note that in deeper layers of optically thick flows, $L_{\rmn{tot}}$ is smaller than the bolometric luminosity of the source due to self-shielding. 

In equation \ref{mdef}, $M$ is the total force multiplier,
\begin{equation}
M=M_{\rmn{bb}}+M_{\rmn{bf}}+M_{\rmn{ff}} +M_{\rmn c},
\label{mparts}
\end{equation}
where $M_{\rmn{bb}}$ is the bound-bound force multiplier,
$M_{\rmn{bf}}$ is the bound-free force multiplier, and $M_{\rmn{ff}}$
is the free-free force multiplier. $M_{\rmn c}$, the Compton force multiplier, equals unity by definition.  

Below we define and discuss each of the terms in equation \ref{mparts}
in the context of a general flow characterized by its number density
profile $n_H(r)$, velocity profile $v(r)$,  and a given thermal and
ionization structure. In all cases we assume a typical type-I AGN
continuum with a ``UV bump'' and a flat X-ray continuum
($\alpha_{ox}=1.4$ and $\alpha_x=0.9$; $L_E \propto E^{-\alpha}$, where $\alpha_x$ is defined over the range 0.1-10\,keV). Solar metallicity gas is assumed throughout.

\subsubsection{Exact and Sobolev-type calculations of $M_{\rmn {bb}}$}

Line absorption is sensitive to differential velocities in excess of the thermal velocity. We use an exact method to compute the radiation pressure force multiplier , $M^{X_i}_l(r)$, for a line $l$ of ion $X^i$ at location $r$, where the velocity is $v(r)$: 
\begin{equation}
\begin{array}{l}
{\displaystyle M^{X_i}_l(r) =\frac{0.015 n^{X_i}f_l
    \lambda_l L_{E_l} }{4\pi r^2 H_c v_{\rmn{th}}}  \times} \\
~~~~~~~~~~~~~{\displaystyle \int_{-\infty}^{\infty} 
 dE' \phi \left [ E'-E_l(1-v(r)/c) \right ] e^{-\tau(E',r)}}, \\
\end{array}
\label{mbb}
\end{equation}
where $n^{X_i}$ is the ion number density, $f_l$ the oscillator
strength, $v_{\rmn{th}}$ the thermal speed for ion $X_i$, $L_{E_l}$ the monochromatic source luminosity at the line energy, $E_l$, and $\phi$ is the line 
profile. We assume pure Doppler line profiles since damping wings have negligible dynamical effects for the densities associated with AGN flows.  The optical depth, $\tau$, at energy $E'=E_l(1-v(r)/c)$ and position $r$ is given by 
\begin{equation}
\begin{array}{l}
{\displaystyle \tau(E',r)=}\\
~~~~~~~~{\displaystyle \sum_{X^i} \left [\int_0^r d\tau_{\rmn
    {bf}}^{X_i}(E',r') + \sum_{l'} \int_0^r d\tau_{l'}^{X_i}(E',r')
\right ],}\\
\end{array}
\label{tau_def}
\end{equation} 
where 
\begin{equation}
\begin{array}{l}
d\tau_{l'}^{X_i}(E',r')=\\
~~~~~~~~~{\displaystyle 0.015n^{X_i}(r')\frac{f_{l'}\lambda_{l'}}{v_{\rmn
    {th}}}\phi(E'-E_{l'}(1-v(r')/c))dr',} \\
\end{array}
\label{tau_l}
\end{equation} 
and $d\tau_{\rmn {bf}}$ is the bound-free optical depth (assumed to vary
little over the line profile). As evident from equations \ref{tau_def} and \ref{tau_l}, the
optical depth depends on the ionization structure and the 
velocity profile up to location $r$. 
This formalism is adequate for calculating the line radiation pressure force including the effect of differential expansion, continuum optical depth, and line interactions (e.g., line blanketing). The total bound-bound force multiplier is given by, 
\begin{equation}
M_{\rmn {bb}}=\sum_{X,i,l}M^{X_i}_l.
\label{mlsum}
\end{equation}
 We refer to this formalism as the exact $M_{\rmn {bb}}$ method.

A simpler and  widely used method is based on the Sobolev approximation. This formalism is outlined in the appendix and assumes that each section of the flow (one Sobolev length across; see appendix) is decoupled from the rest of the flow in terms of its line optical depth.  A useful definition  is the local optical depth for Thomson scattering (e.g., Arav, Li, \& Begelman 1994),
\begin{equation}
t=n_e \sigma_T v_{\rmn s} \left ( \frac{dv}{dr} \right )^{-1},
\label{t_sobol}
\end{equation}
where $v_s$ is the sound speed. The Sobolev line force multiplier is thus a function of the gas ionization level and $t$ (see appendix), which are both local quantities. This dependence on local properties is the great advantage of the Sobolev method over exact radiation transfer methods since the calculation of the radiation pressure force is less time consuming. Below we compare the exact and the Sobolev methods for various physical situations.

Previous works in stellar flows have shown that, for LTE conditions,
some $10^5$ lines are required to fully account for $M_{\rmn {bb}}$ (e.g., Stevens \& Kallman 1991). However, the low density AGN flows are very different. Our NLTE calculations show that the population of  most excited levels is small, and that  all ions are effectively in their ground level. For this reason, only resonance and inner shell lines contribute to $M_{\rmn {bb}}$. Our code includes some 2000 resonance and inner shell lines compiled from Verner, Verner, \& Ferland (1993), Behar et al. (2001), and Behar \& Netzer (2002). In the NLTE conditions under study, collisional ionization is not very important, and the thermal-ionization structure is determined almost completely by the shape of the ionizing continuum and the ionization parameter, $U_x$. The ionization parameter is defined here as the ratio of the photon number density in the energy range  0.1-10\,keV to the hydrogen number density. In this notation, broad line region (BLR) clouds are characterized by $U_x\ltorder10^{-3}$ while the highly ionized X-ray gas is at $U_x\simeq 0.1$ (CN01 and references therein). 

Stevens \& Kallman (1991) find that the most important lines
contributing to $M$ are optical and near--UV lines (see their table
1). The main difference between their work and ours results from their
LTE assumption where the excited states are populated and higher
transitions contribute significantly to $M_{\rmn {bb}}$. In AGN, the
largest contribution is due to lines with energies around the UV bump
(10--50\,eV; see table \ref{t:lines}). This is true for a wide range
of $U_x$. Thus, the gas dynamics in AGN is sensitive to the luminosity
of the UV bump which is not directly observed. Our calculations show
that only a few lines are needed to account for most of the line
radiation pressure for a wide range of $U_x$ and $t$. Specifically,
90\% of $M_{\rmn {bb}}$ in optically thin AGN gas are due to the ten
strongest lines. As $t$ increases, more lines are contributing
significantly to $M_{\rmn {bb}}$ (e.g., some 100 lines are needed to
account for 90\% of $M_{\rmn {bb}}$ for $U_x=10^{-3}$ and
$t=10^{-4}$). For extremely large $t$, $M_{\rmn {bb}} \ll M_{\rmn {bf}}$. 

\begin{table} 
\centering 
\begin{tabular}{lcc} 
\hline 
Line      &    $\tau_l$ &  $M_l$ \\ 
\hline 
& $U_x=10^{-3}$ & ($M_{\rmn {bb}}=4$)\\
\hline
H{\sc i} ($\lambda 951$)  &   8     &  0.1        \\
H{\sc i} ($\lambda 973$)  &   17      &  0.08        \\
H{\sc i} ($\lambda 1027$)  &    50     &  0.07        \\
C{\sc iv} ($\lambda 1549$)  &    450     &  0.01        \\
\hline
& $U_x=10^{-2}$ & ($M_{\rmn {bb}}=3$) \\
\hline
H{\sc i} ($\lambda 973$)  &   1.5      &  0.16        \\
He{\sc ii} ($\lambda 1027$)  &   4      &  0.13        \\
He{\sc ii} ($\lambda 951$)  &   0.8      &  0.12        \\
C{\sc iv} ($\lambda 1549$)  &    27     &  0.02        \\
\hline
& $U_x=10^{-1}$ & ($M_{\rmn {bb}}=3$)\\
\hline
He{\sc ii} ($\lambda 256$)  &   0.8      &  0.18        \\
H{\sc i} ($\lambda 1216$)  &   0.8      &  0.16        \\
Ne{\sc vi} ($\lambda 402$)  &   2      &  0.15        \\
C{\sc iv} ($\lambda 1549$)  &    0.05     &  0.05        \\
\hline
& $U_x=1$ & ($M_{\rmn {bb}}=1$) \\
\hline
S{\sc xiv} ($\lambda 427$)  &   0.9      &  0.14        \\
Si{\sc xii} ($\lambda 368$)  &   0.8      &  0.12        \\
Ar{\sc xvi} ($\lambda 303.5$)  &   0.3      &  0.07        \\
C{\sc iv} ($\lambda 1549$)  &    0     &  0        \\
\hline
& $U_x=10$ & ($M_{\rmn {bb}}=0.4$)\\
\hline
Fe{\sc xxiv} ($\lambda 192$)  &   0.7      &  0.14        \\
Fe{\sc xxiv} ($\lambda 255$)  &   0.3      &  0.09        \\
Fe{\sc xxiii} ($\lambda 133$)  &  1.0      &  0.04        \\
C{\sc iv} ($\lambda 1549$)  &    0     &  0        \\
\hline
\end{tabular}
\caption{Contribution of the most important lines to the force
  multiplier assuming a Sobolev $t$ parameter of $10^{-4}$ (see text and compare with table 1 in Stevens \& Kallman 1991). Also shown is the bound-bound force multiplier, $M_{\rmn {bb}}$, for $t=10^{-4}$ neglecting line blanketing. Note that in all cases, lines whose energies are near the UV bump contribute most to the radiation pressure force. The contribution of \civ~$\lambda1549$ to $M_{\rmn {bb}}$ is shown for all cases.} 
\label{t:lines}
\end{table}

\subsubsection{Continuum absorption}

The bound-free force multiplier is
\begin{equation}
M_{\rmn {bf}}(r)=\sum_{X,i} \frac{n^{X_i}}{n_e \sigma_T L_{\rmn
    {tot}}} \int_0^\infty  \sigma_E^{X^i}(E) L_E(E)e^{-\tau(E,r)}dE,
\label{mbf}
\end{equation}
where $\sigma(E)$ is the bound-free cross-section from all levels (including inner shells) and $\tau(E,r)$ is defined by equation \ref{tau_def}. As shown by Arav, Li, \& Begelman (1994), and in CN01, the contribution of $M_{\rmn {bf}}$ to $M$ cannot be neglected for both optically thick and optically thin flows. The reason is (e.g., Netzer 1996) that the integrated cross section of a bound-free edge ($\int \sigma_EdE$) is comparable to that of the strongest resonance transition of the same series. Thus, for optically thin gas, for ion $X_i$, 
\begin{equation}
\frac{M^{X_i}_{\rmn {bb}}}{M^{X_i}_{\rmn {bf}}} \simeq 
\frac{(E_l/E_0)^{-\alpha}(\pi
  e^2/mc)f_{l}}{\int_{E_0}^\infty
 (E/E_0)^{-\alpha} \sigma_E dE} \simeq \frac{2+\alpha}{\left
    ( E_l/E_0 \right )^{\alpha_{E_l E_0}}},
\label{ratio_bb_bf}
\end{equation}
where $\alpha_{E_l E_0}$ is the continuum slope between the line energy
and the edge threshold energy, and $\alpha$ is the continuum slope above the edge. For our typical type-I AGN continuum, a good approximation in the range $10^{-6}<U_x<10$ is $M_{\rmn {bf}}/M_{\rmn {bb}} \simeq 0.04U_x^{-0.17}$. 

The contribution of free-free processes can be neglected for flow densities $<10^{12}~{\rmn {cm}^{-3}}$ (CN01) and typical infra-red luminosities of type-I AGN.

\subsection{The internal radiation pressure}

\begin{figure}
\includegraphics[width=84mm]{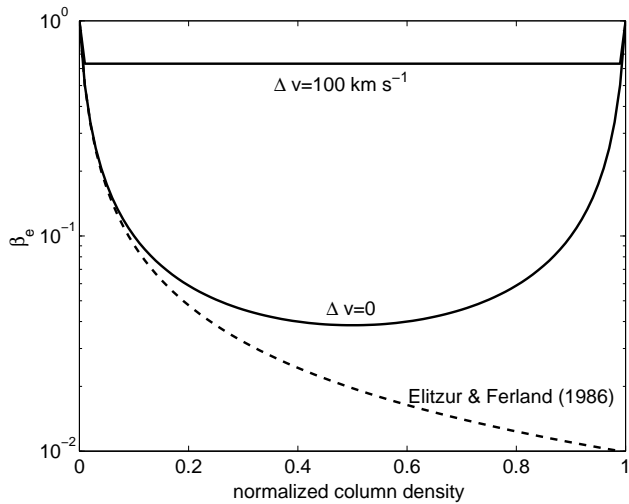}
\caption{The escape probability, $\beta_e$, as defined by EF86 (dashed curve), compared to the one defined in equation \ref{beta} (solid lines) for static ($\Delta v=0$) and differentially expanding slabs with a uniform velocity gradient. In both cases $v_{\rmn {th}}=10~{\rmn {km~s}^{-1}}$ and the total (static) line optical depth is 100. Note the steep increase in $\beta_e$ towards the boundaries. Also note the dramatic effect of differential expansion on $\beta_e$ (the $\Delta v=100~{\rmn {km~s}^{-1}}$ line). In this case, the line optical depth is determined by  the velocity gradient rather than the column density (see text).}
\label{beta_e}
\end{figure}

The total pressure inside the flow is the sum of the gas pressure, $P_{\rmn {gas}}$, and the isotropic radiation pressure, $P_{\rmn {rad}}$, which results from the scattering of resonance line photons. As shown by EF86, in BLR clouds $P_{\rmn {rad}}$ is of the order of $P_{\rmn {gas}}$. The numerical scheme provided by EF86 takes into account semi-infinite, uniform slabs. According to this scheme,
\begin{equation}
\frac{P_{\rmn {rad}}}{P_{\rmn {gas}}}=\sum_l \frac{8\pi \nu^3 v_{\rmn {th}}}{3 n_e c^4}
  \frac{h \nu}{k_BT_x}  \frac{1-\beta_e}{ e^{h\nu /k_B T_x}-1},
\label{pres_mov}
\end{equation}
where $T_x$ is the excitation temperature which is
determined from the non-LTE calculation and $\beta_e$ the
local escape probability. AGN flows are stratified, dynamically expanding, and have finite total optical depth. Hence, $\beta_e$ must be modified to account for these conditions. A good approximation, within the framework of the escape probability formalism, is
\begin{equation}
\beta_e={\rmn {max}}(\frac{1}{1+\tau_l},\frac{1-e^{-\tau_s}}{\tau_s}),
\label{beta}
\end{equation}
where 
\begin{equation}
\tau_l \equiv \left (\frac{1}{\tau_l^+}+\frac{1}{\tau_l^-} \right )^{-1},
\label{taupm}
\end{equation} 
$\tau_l^\pm(r)$ denotes the optical depth at location $r$ to the two boundaries and $\tau_s$ is the Sobolev line optical depth (see equation A1). This expression is valid provided the gas ionization level changes monotonically and non-local scattering of emission line photons can be neglected. Figure \ref{beta_e} shows the effect of the finite size of the flow on $\beta_e$, where close to both boundaries photons escape the gas. Dynamics has a dramatic effect on $\beta_e$, as line optical depths are reduced, and photons can escape even from the innermost layers of the flow. The deviation from the simplified EF86 method, which assumes $\beta_e \simeq (1+\tau_l^+)^{-1}$, can amount to several orders of magnitude in $\beta_e$.
The next section contains a detailed example of the distribution of $P_{\rmn {rad}}/P_{\rmn {gas}}$ for optically thin and optically thick uniform flows.

\subsection{The numerical scheme}

We have developed a numerical procedure for calculating the radiation
pressure force and the global self-consistent structure and dynamics
of AGN flows. For given flow properties, $v(r),~n(r)$, and $L_{\rmn
  {tot}}$, we calculate the ionization and thermal structure of the
flowing gas under NLTE conditions, using {\sc ion2002}, the 2002
version of the photoionization code {\sc ion} (Netzer 1996). Given the
ionization structure, we construct an optical depth vs. velocity  grid
for every line. The velocity resolution  is $0.3 v_{\rmn {th}}$ of the
respective ion. We divide the flow into thin layers with velocity
steps of $0.3 v_{\rmn {th}}$ of the heaviest ion. A finer local grid is
used in cases where the gas properties change on smaller scales (e.g.,
ionization fronts). For each layer, we calculate the corresponding
optical depth grids, taking into account the local velocity and the
continuum opacity, and obtain $M_{\rmn {bb}}$, $M_{\rmn {bf}}$ and $H_c$.
Once these quantities are calculated, $F_{\rmn {rad}}$ is obtained and
equation \ref{eqnmot} can be solved. The code optimizes for redundant
lines whose contribution is negligible either due to the low abundance
of the relevant ions or due to negligible flux levels at the line energy. Given the radiation pressure force, we can move on to calculate the self-consistent flow structure and dynamics. 

Our calculations of the radiation pressure force are considerably
different from those used in previous stellar-type works in the sense
that they do not depend explicitly on $dv/dr$. Therefore the equation
of motion does not have the usual form ($r^2vdv/dr \propto
(r^2vdv/dr)^\alpha$; see appendix A) and does not necessarily give the
standard wind solution $v\propto (1-r/r_0)^{1/2}$. As shown by Poe,
Owocki, \& Castor (1990) for the case of pure absorption, isothermal flows, non-Sobolev solutions are considerably different from Sobolev solutions in terms of their dynamical stability and the lack of critical points apart from the sonic point. Our model is different in several other ways, e.g., we do not assume a constant thermal width for all lines and consider ionization and temperature gradients across the flow.  As we show below, the radiation pressure force depends also on the velocity of the flow, and is therefore very different from that used by Poe, Owocki, \& Castor (1990). These differences are added to the inherent differences between AGN and stellar flows due to the different geometry. Another major difference is that stellar winds start subsonically, while the absorption troughs of AGN flows imply flows which are supersonic throughout. 

In our model, the flow is characterized by its launching point $r_0$, the density $n_H(r_0)$, and the initial velocity $v_0$. Given the AGN luminosity, $L_{\rmn {tot}}$, and the spectral energy distribution (SED), we determine the ionization parameter, $U_x(r_0)$ at the base of the flow. Gravity is expressed using the Eddington ratio, $\Gamma$, where $g \equiv H_c/\rho \Gamma$ (see CN01). The first iteration assumes a constant velocity, $v=v_0$, and $\rho \propto r^{-2}$. We then calculate the radiation pressure force, solve equation \ref{eqnmot} and obtain $v(r)$. A new density profile, $n_H(r)$, is obtained from the continuity condition. $F_{\rmn {rad}}$ is then recalculated and the next iteration starts. The process is repeated  until 1\% convergence of the flow density profile is achieved. Typically, the code takes several hours to run on a PC linux machine for high velocity ($\sim 10^4~{\rmn{km~s}^{-1}}$) stratified flows. 

The scheme presented in this paper assumes $v_0=v_s$, i.e., initially transonic flows, and is very different from the method used in several previous works where the mass flow rate was determined from critical point analyses. In a non-Sobolev calculation, the only critical point is the sonic point (e.g., Abbott 1982). In AGN flows, the sound speed is determined by photoionization and is not related to the Keplerian velocity of the gas. The flow is supersonic throughout and the mass flow rate is determined solely by the initial conditions. More details are given in appendix B.

\section{Results}

In this section we investigate the differences  between the exact method and the Sobolev approximation. We first study the radiation pressure force distribution in uniform density clouds and then generalize the results to  self-consistent BAL and NAL flows. The radiation pressure distribution inside optically thin and thick flows is also discussed. We choose cases where hydrogen and helium are fully ionized in order to allow a meaningful comparison between the Sobolev and the exact schemes.

\subsection{Exact and Sobolev line profiles} 

\begin{figure}
\includegraphics[width=84mm]{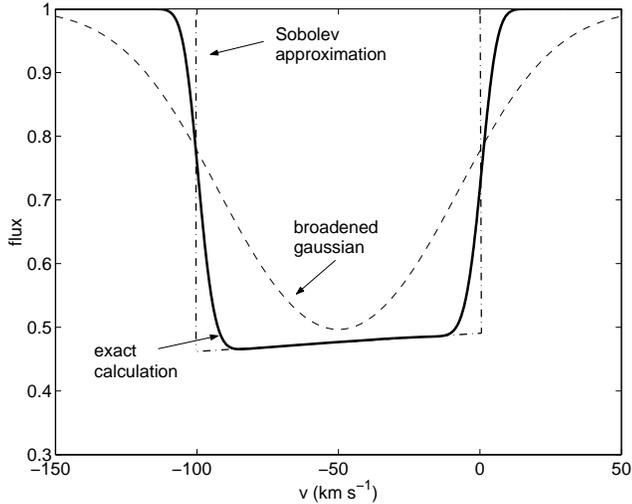} 
\caption{An exact absorption line profile (solid line) and a Sobolev line profile (dash-dotted line) for a flow with $\Delta v=100~{\rmn {km~s}^{-1}}$ and $N_H=10^{22}~{\rmn {cm}^{-2}}$. A broadened Gaussian line profile with FWHM=$100~{\rmn {km~s}^{-1}}$ is shown for comparison (dashed line). The Sobolev method produces a wrong profile at both low and high velocities. Gaussian line profiles provide an even worse approximation to the true profile. The deviation from a flat profile in both the exact and the Sobolev method is due to small changes in the ionization structure (see also section 3.4).}
\label{lprofs}
\end{figure}

The fundamental difference between the exact method and the Sobolev approximation lies in their definition of the line optical depth. This is illustrated in the numerical example shown in figure \ref{lprofs} corresponding to a slab of thickness $\Delta R$, a column density, $N_H=10^{22}~{\rmn {cm}^{-2}}$, and velocity interval $\Delta v=100~{\rmn {km~s}^{-1}}$ with a constant $dv/dr$ across the slab.  The resulting  line profiles are very different. The Sobolev approximation fails to predict the line wings since it assumes a constant optical depth across one Sobolev length and thus a local, rectangular line profile. The line wings predicted by the exact calculations are reminiscent of the Doppler line profile which characterizes the local absorption cross-section in the flow. For comparison, we also show a broadened Gaussian line profile with FWHM=$100~{\rmn {km~s}^{-1}}$. Such line profiles are sometime assumed in order to calculate the radiation pressure force. This profile is different from both the exact and the Sobolev profiles. The exact calculation shows that the contribution of the line wings to the optical depth is substantial. Such line profiles cannot be adequately modeled by the Sobolev approximation whose {\it local} line profile is infinitesimally narrow. Such inaccuracies can significantly alter the radiation pressure force distribution inside a flow, and hence the flow dynamics.

\subsection{Exact and Sobolev force multipliers}

\begin{figure}
\includegraphics[width=84mm]{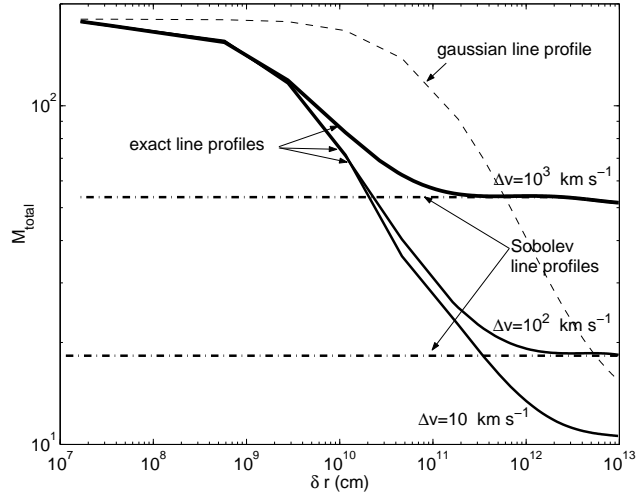}
\caption{The total force multiplier calculated with the exact formalism (solid line), the Sobolev approximation (dot-dashed line), and assuming broadened Gaussian lines (dashed line). The Sobolev method is a poor approximation for $M$ at the illuminated surface but converges to the exact result deeper in the cloud. Broadened Gaussian lines (shown for a micro-turbulence velocity of $100~{\rmn {km~s}^{-1}}$) provide a poor approximation for $M$.}
\label{sharp}
\end{figure}

Figure \ref{sharp} shows the radiation pressure force distribution for a cloud with $U_x=10^{-2.5}$ and $N_H=10^{22}~{\rmn {cm}^{-2}}$ for various values of the velocity gradients $\Delta v/\Delta R$. For $\Delta v =10~{\rmn {km~s}^{-1}}$, the solution is almost identical to the static solution (see e.g., CN01), where $M$ monotonically declines into the gas due to the increasing optical depth. For larger $\Delta v$, $M$ declines from the illuminated surface and then flattens as some $\delta r<\Delta R$. This results from the fact that, up to some location, the gas differential velocity from its illuminated surface is smaller than $\left <v_{\rmn {th}} \right >$ and the gas can be considered static. Here $\left < v_{\rmn {th}}\right >$ is an average over the most important ions which drive the gas, and is usually smaller than the sound speed. At larger $\delta r$, the flow velocity exceeds $\left <v_{\rmn {th}} \right >$ and  the optical depth saturates. For a uniform cloud this results in a constant optical depth per unit velocity, and therefore constant $M$, which is basically the Sobolev $M$.  In all cases and for all $\Delta v$, the exact force multiplier exceeds the Sobolev force multiplier at the illuminated surface by a large factor (figure \ref{sharp}). This is a general behaviour when sharp discontinuous flow surfaces are present. We shall refer to this  as the ``skin effect''. The reason is that the Sobolev approximation is insensitive to length scales smaller than the Sobolev length (see appendix). For AGN flows, the column density  of the skin layer is
\begin{equation}
N_H^s \sim {\rmn {min}} \left ( n_H \left <v_{\rmn {th}} \right > \frac{dr}{dv}, 10^{21}U_x\right )~{\rmn {cm}^{-2}}.
\label{sobad}
\end{equation}
The first term  depends on $\left <v_{\rmn {th}} \right >$ which can be set to $\sim 13~{\rmn {km~s}^{-1}}$ (the sound speed at $T_e=10^4$K) and is valid for a wide range of ionization parameters since at much higher temperatures, the highly ionized gas is driven by heavy metals whose thermal line width is small. The second term in this equation reflects the fact that for $N_H\gtorder 10^{21}U_x~{\rmn {cm}^{-2}}$, the contribution of lines to the force multiplier in the subsonic part of the flow can be neglected. To conclude, flows with rapidly varying properties over small scales, cannot be modeled properly by the Sobolev approximation, and the best approximation for $M$ is obtained by assuming static gas.

Comparison with the exact and Sobolev methods, we also show the effect of broadened Gaussian lines on the distribution of $M$ (figure \ref{sharp}).  Gaussian models overpredict $M$ close to the illuminated surface and under-predict it for the deeper layers. 

\begin{figure}
\includegraphics[width=84mm]{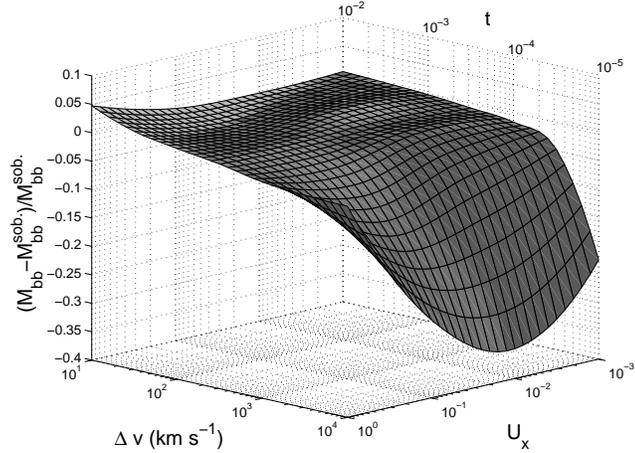} 
\caption{The relative difference between the line force multiplier computed with the Sobolev approximation and the exact numerical calculations.  The calculations assume constant ionization structure across a $10^{22}~{\rmn {cm}^{-2}}$ flow. For the faster moving flows, line troughs overlap, line blanketing becomes important and the exact method deviates more from the Sobolev approximation. Line blanketing becomes important with increasing $\Delta v$ since more lines overlap and $M_{\rmn {bb}}$ contributes a larger fraction to $M$.}
\label{blank}
\end{figure}

While the Sobolev approximation agrees with the exact calculations beyond $\delta v\equiv v(r_0+\delta r)-v(r_0)\sim \left < v_{\rmn {th}} \right >$, it cannot account for line interaction, such as line blanketing.  The effect of line interaction is to reduce the force multiplier, hence the calculated $M_{\rmn {bb}}^{\rmn {sob}.}$ which does not include this effect, is artificially large. Figure \ref{blank} shows the effect of line blanketing on $M_{\rmn {bb}}$ for a constant density cloud with $N_H=10^{22}~{\rmn {cm}^{-2}}$ and various values of $\Delta v$. We neglect the skin effect and compare the line force multiplier towards the non-illuminated part of the cloud.  The effect of line blanketing is small below $\Delta v \sim 3000~{\rmn {km~s}^{-1}}$ but becomes significant and reduces $M$ by a large factor at large velocities over almost the entire range of $U_x$ (except for extremely ionized gas where $M_{\rmn {bb}} \ll 1$). This is due to the fact that as the velocity range increases, more lines are able to interact. Furthermore, as $\Delta v$ increases, $M_{\rmn {bb}}$ contributes more to $M$, and the effect on the gas dynamics is more pronounced. The effect of line blanketing is non-local, and as such, depends on the {\it global} ionization structure of the flow. For this reason, it must be solved self-consistently for each flow model (see section 3.2). Our results are in qualitative agreement with those of Abbott (1982) who investigated line blanketing in stellar atmospheres.

\begin{figure}
\includegraphics[width=84mm]{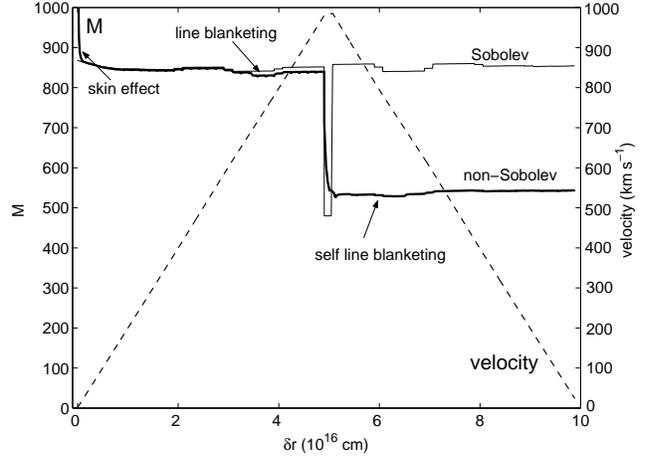} 
\caption{The effect of self-blanketing on $M$ in non-monotonic flows. Notice the large discrepancy between the Sobolev method and the exact calculations due to non-local line interaction for $\Delta r>5\times10^{16}~{\rmn {cm}}$. Note also the ``skin effect'' close to $\delta r\simeq 0$, and the small but noticeable effect of line blanketing for $3<\delta r <5\times10^{16}~{\rmn {cm}}$.}
\label{selfb}
\end{figure}

An important  case is self-blanketing which occurs in non-monotonic flows. We illustrate this by choosing a generic model where a uniform density flow, with a column density of $10^{21}~{\rmn {cm}^{-2}}$, is initially accelerated with a constant velocity gradient $2\Delta v/\Delta R$, and then decelerated with the same gradient (dashed line in figure \ref{selfb}). The exact calculations clearly show  a sharp drop in $M$ beyond the deceleration point since lines trace back their absorption profiles and are exposed to lower flux levels. The Sobolev method, that is only defined locally, cannot account for such effects and overestimates M beyond this point. It also fails to correctly predict $M_{\rmn {bb}}$ for $dv/dr=0$  since at this point, the Sobolev optical depth diverges and $M_{\rmn {bb}}=0$. Also marked in figure \ref{selfb} are the (small) effects of blanketing by other lines, and the skin effect close to the illuminated surface ($\delta r =0$). 

The locking of absorption lines onto emission lines is another potentially important effect (Korista et al. 1996; Arav 1996). Here we consider two kinds of line locking: self line locking where resonance absorption lines are exposed to an enhanced flux level due to emission by the same line (e.g., $L_\alpha-L_\alpha$ locking), and line locking by different lines. The best known example of the second kind is the ``$L_\alpha$ ghost'' phenomenon in BALQSOs (e.g., Arav 1996).

\begin{figure}
\includegraphics[width=84mm]{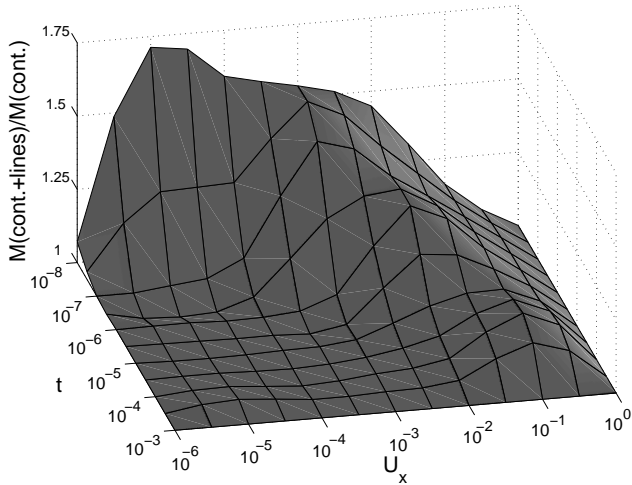} 
\caption{The effect of self line locking on the total (line and continuum) radiation pressure force. The ratio of the radiation pressure force when locking is taken into account, to the case where it is neglected is shown as a function of $U_x$ and $t$. Note that line locking is most pronounced for large velocity gradients (small $t$) since lines are optically thin. For moderately ionized gas the effect is pronounced for larger $t$. The radiation pressure force in highly ionized gas ($U_x>0.1$) is not affected by line locking on optical and UV emission lines.}
\label{mlock}
\end{figure}

We first consider self line locking. All emission line properties are taken from the composite spectrum of Telfer et al. (2002). The line list include \mgii~$\lambda 2798$, \civ~$\lambda 1549$, $L_\alpha$, \nv~$\lambda 1240$, and \ovi~$\lambda 1035$ (the contribution of other lines is expected to be small for BLR gas) and a constant full width at half maximum, FWHM=$4000~{\rmn {km~s}^{-1}}$ was assumed. This results in \hi~ $L_\alpha$ peak flux level being 5 times that of the adjacent continuum. In figure \ref{mlock} we show the ratio of radiation pressure force with and without line locking. For low ionization flows ($U_x \ltorder 10^{-4}$), the radiation pressure force is larger by a factor $\gtorder 2$ when line locking is taken into account. This is mainly due to the locking of $L_\alpha$ absorption on $L_{\alpha}$ emission. For larger $t$,  the contribution of $L_\alpha$ to the force multiplier drops since the line becomes optically thick.  For moderate ionization ($10^{-4}<U_x<10^{-2}$), locking by higher ionization lines becomes important (e.g., \civ~$\lambda 1549$). As $U_x$ increases, the effect is smaller but extends to larger $t$. This results from lines being more optically thin for larger $t$.   For highly ionized gas ($U_x>0.1$) the contribution of optical and near-UV lines to $M_{\rmn {bb}}$ is negligible. Obviously, the above results depend somewhat on the assumed emission line properties.

We have examined the \nv~$\lambda 1240-{\rmn L}_\alpha$ interaction and its effect on the radiation pressure force by computing the ratio of the radiation pressure force due to \nv~$\lambda 1240$ ($M^{\mbox{\nv}}_{\lambda 1240}$) on $M_{\rmn {bb}}$ for various uniform density cloud models. When the flow velocity exceeds $3000~{\rmn {km~s}^{-1}}$ (depending on the FWHM of $L_\alpha$), \nv~$\lambda 1240$ is exposed to higher flux levels and $M^{\mbox{\nv}}_{\lambda 1240}$ rises. Once $v\gtorder 5900~{\rmn {km~s}^{-1}}$, $M^{\mbox{\nv}}_{\lambda 1240}$ drops due to blanketing with $L_\alpha$ absorption line at low velocities. Our results show that the maximum relative contribution of \nv~$\lambda 1240$ to $M$ is can be approximated by 
\begin{equation}
\frac{M_{\lambda 1240}^{\mbox{\nv}}}{M} \simeq 0.03 \left ( \frac{\rmn {N/H}}{10^{-4}} \right ) \left ( \frac{\rmn {4000~ km~s}^{-1}}{\rmn {FWHM(L}_\alpha)} \right ) \left ( \frac{\rmn {EW(L}_\alpha)}{100\mbox{\AA}} \right ).
\label{nvla}
\end{equation}
This implies that, for the assumed SED and gas metallicity, the dynamical effect of  \nv~$\lambda 1240$ locking onto $L_\alpha$ is small. As the radiation pressure force is roughly proportional to the gas metallicity, the effect is larger for overabundant nitrogen. It is also more significant when the ionizing flux is affected by shielding as will be discussed in a forthcoming paper.


\subsection{A comparison of the internal line radiation pressure}

\begin{figure}
\includegraphics[width=84mm]{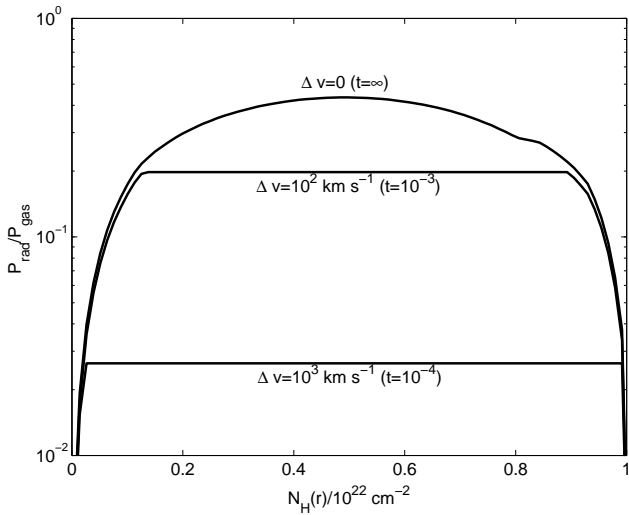} 
\caption{Internal radiation pressure to gas pressure ratio for a typical highly ionized gas cloud with a colum density of $10^{22}~{\rmn {cm}^{-2}}$. For static gas ($\Delta v=0$), this ratio reaches a maximum  at the  center of the cloud, and declines to zero towards the  non-illuminated boundary. For a finite $\Delta v$, a  constant ratio is maintained across a large section of the cloud, and the peak $P_{\rmn {rad}}/P_{\rmn {gas}}$ is lower.}
\label{prad2pgas}
\end{figure}

As explained in section 2, the distribution of the internal radiation pressure can affect the flow structure. Some examples for static and expanding uniform  clouds are shown in figure \ref{prad2pgas}. Our results differ significantly from those of EF86 since in our modified procedure (equations \ref{pres_mov}, \& \ref{beta}), line photons  escape the cloud through both (all) boundaries. This suppresses $P_{\rmn {rad}}$ close to the cloud's rims. Introducing a constant velocity gradient decreases the line optical depth (figure \ref{beta_e}) and, for uniform clouds, maintains a constant value of $P_{\rmn {rad}}$ across most of the flow. This is shown in figure \ref{prad2pgas} for $t=\infty$, $t=10^{-3}$, and $t=10^{-4}$. Larger velocity gradients result in smaller $P_{\rmn {rad}}$. Thus, differentially expanding clouds are less disrupted by $P_{\rmn {rad}}$. For optically thin clouds, the peak $P_{\rmn {rad}}/P_{\rmn {gas}}$ is at the cloud centre. 

\begin{figure}
\includegraphics[width=84mm]{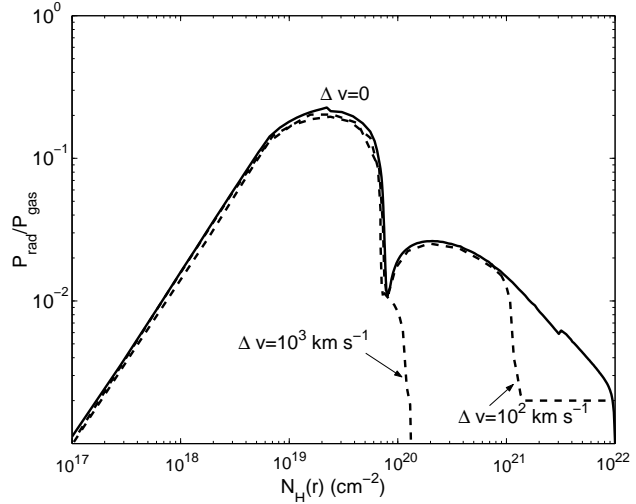}
\caption{Radiation pressure to gas pressure ratio for a gas with $U_x=10^{-3}$ and $N_H=10^{22}~{\rmn {cm}^{-2}}$. The profile is non-monotonic due to the varying ionization structure. The left hump is mainly due to the scattering of \civ~$\lambda 1549$ photons and the right hump due to \heii~$\lambda 304$ line photons.  $P_{\rmn {rad}}$ is suppressed once $\delta v \gtorder v_s$ (note the forking of the dashed curves from the solid line for $\Delta v=10^2,~10^3~{\rmn {km~s}^{-1}}$).} 
\label{prad_thick}
\end{figure}

The above discussion applied to gas that is optically thin to continuum absorption. Optically thick clouds show a more intricate $P_{\rmn {rad}}/P_{\rmn {gas}}$ profile, as can be seen from figure \ref{prad_thick}. Such profiles have, in general, several maxima (two in this case) due to the varying ionization structure. In the case shown, the first hump is mainly due to the scattering of \civ~$\lambda 1549$ line photons. When self-shielding becomes substantial, and the ionization structure changes significantly, ($N_H \gtorder 10^{20}~{\rmn {cm}^{-2}}$ in figure \ref{prad_thick}), the outward optical depth in lines of highly ionized species (\civ~$\lambda 1549$ in this example) drops, and line photons do not contribute to $P_{\rmn {rad}}$ any more. The second hump is due to scattering in lines of lower ionization species (in this case mostly \heii~$\lambda 304$). Velocity gradients (see the $\Delta v=100,~1000~{\rmn {km~s}^{-1}}$ curves) suppress $P_{\rmn {rad}}$ once $\delta v$  exceeds the thermal velocity of the ions whose lines contribute most to $P_{\rmn {rad}}$. For smaller $\delta v$, the gas is essentially static as far as optical depth effects are concerned and $P_{\rmn {rad}}$ is similar to the static case. 

\subsection{Self-consistent flow calculations}

In this section we study the self-consistent structure and dynamics of AGN flows using the exact and the Sobolev methods. Comparing the two, we are able to explore the effect of line locking and blanketing. We also discuss the dynamical implications of the skin effect.

\subsubsection{A linear approximation}

To assess the dynamical effect of line blanketing and locking we write the equation of motion for the supersonic section of the flow as
\begin{equation}
v\frac{dv}{dr}=\frac{H_c}{\rho} \left ( M+\delta M \right ),
\label{eqnmotp}
\end{equation}
where $\delta M$ is a small perturbation ($\vert \delta M/M \vert < 1$) representing the effect of line blanketing and locking. As a first order perturbative analysis we use the unperturbed velocity profile, $\tilde{v}(r)$, to express the ratio of the terminal velocities with and without the perturbation,
\begin{equation}
\frac{v_\infty}{\tilde{v}_\infty} \simeq 1+\left ( \frac{1}{\tilde{v}_\infty} \right )^2\int_{r_0}^\infty \frac{H_c}{\rho}\delta M(\tilde{v},\tilde{v}d\tilde{v}/dr)dr,
\label{vrat}
\end{equation}
where $\tilde{v}_\infty$ refers to the unperturbed terminal velocity (the mass flow rate is determined by the initial conditions of the flow  and not from critical point analyses which justifies the use of the present perturbation analysis).

The effect of line locking can be written as
\begin{equation}
\Delta M=(A-1)M\phi_e(\tilde{v},v_e),
\label{mlock1}
\end{equation}
where $\phi_e$ is the emission line profile and $v_e$  the characteristic line width ($\simeq 0.5$\,FWHM). We assume a Doppler emission line profile $\phi_e\propto e^{-\tilde{v}^2/v_e^2}$ and  $A=A(U_x,t)$ is the factor by which the force multiplier increases due to locking and is  given in figure \ref{mlock}. A Sobolev treatment yields the canonical wind solution: $\tilde{v}d\tilde{v}/dr=$const, i.e.,  $\tilde{v}=\tilde{v}_\infty (1-r_0/r)^{1/2}$. Assuming that $U_x=$const across the flow (see e.g., the BAL models of Arav, Li ,\& Begelman 1994) then $A=$const, and the first order correction to the terminal velocity is,
\begin{equation}
\frac{v_\infty}{\tilde{v}_\infty} \simeq 1+(A-1) \left(\frac{v_e}{\tilde{v}_\infty} \right )^2. 
\label{vlock}
\end{equation}
This result is insensitive to the exact emission line profile provided
that its wings decline as fast or faster than $v^{-2}$. For an ionized flow with $\tilde{v}_\infty\sim 1000~{\rmn {km~s}^{-1}}$, $v_\infty$ increases by $\sim 5\%$ due to line locking. This approximation is adequate for flows whose line optical depth are less that unity. It provides  a lower limit when lines are optically thick since non-linear effects (e.g., due to the increase in $M$ and $A$) become important.

The effect of line blanketing is non-local by definition. For simplicity we assume it becomes important beyond some velocity $v_b$, and cause a decrease in $M$  by a factor $B$.  In this case,
\begin{equation}
\Delta M=-\Theta(v_b)(1-B)M,
\label{mblank}
\end{equation}
where $\Theta(v_b)$ is the Heaviside function ($\Theta(v\geq v_b)=1,~\Theta(v<v_b)=0$). We thus obtain,
\begin{equation}
\frac{v_\infty}{\tilde{v}_\infty} \simeq  1-B  \left (1-\frac{v_b^2}{\tilde{v}_\infty^2} \right )~~~~{\rmn {for}}~ \tilde{v}_\infty>v_b.
\label{vblank}
\end{equation}
As expected, line blanketing reduces $v_\infty$ and is more important
for high velocity flows.  For example, an ionized, high velocity ($\tilde{v}_\infty\sim10^4~{\rmn {km~s}^{-1}}$) flow for which $B\sim0.2$ (see figure \ref{blank}) results in a terminal velocity which is $\sim10\%$ smaller. 

\subsubsection{Numerical calculations}

We now study the effect of line blanketing and locking on more
realistic flows whose structure, dynamics and spectral features are
calculated, self-consistently, using the algorithm presented in section
2.4. We adopt the model proposed by Arav, Li, \& Begelman (1994)
where the flow consists of small clouds and occupies only a small
fraction, $\epsilon$, of the volume. The continuity condition is,
\begin{equation}
\rho \epsilon r^2 v ={\rmn {const}},
\label{acont}
\end{equation}
 where $\rho \propto r^{-2}$ and $\epsilon \propto v^{-1}$.  As
 explained earlier, we consider supersonic flows  and iterate until the filling factor profile, $\epsilon(r,v)$, converges to $<1\%$. In the present calculations
the flow is illuminated by the unobscured  AGN
continuum and the effects of external shielding will be discussed in a
following paper. For illustrating the result we chose $L_{\rmn
  {tot}}=10^{45}~{\rmn {erg~s}^{-1}}, ~ r_0=10^{18}~{\rmn
  {cm}},~U_x=10^{-2.5}$, and $\epsilon(r_0)=10^{-3.5}$. We also assume
that gravity and pressure gradient force  can be neglected. 

\begin{figure}
\includegraphics[width=84mm]{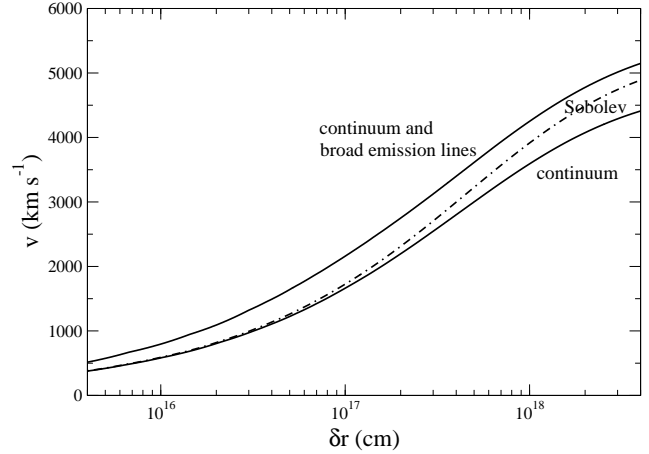} 
\caption{Self-consistent velocity profiles for flows which are driven
  a) by continuum and broad emission lines and b) by continuum
  only. The Sobolev solution for the first scenario is shown for
  comparison (dash-dotted line). Note that line locking on emission lines is dynamically important and increases the flow terminal velocity by $\sim 20\%$. The effect of line blanketing is also important and tends to flatten the velocity profile at high velocities compared to the Sobolev solution.}
\label{velos}
\end{figure}

Figure \ref{velos} shows a comparison of velocity profiles for flows
exposed to continuum with and without emission lines (see section
3.2). $U_x$, $L_{\rmn {tot}}$, and $r_0$ are the same in both cases.  As
shown, line locking increases the final velocity by $\sim 20\%$, which
is in agreement with equation \ref{vlock}. This results from the fact
that, in this model, \ovi~$\lambda 1035$, $L_\alpha$, \nv~$\lambda
1240$, and  \civ~$\lambda 1549$ all contribute significantly to the
force multiplier. We note, however, that the relative importance of
line locking depends on the flux in the UV bump, where many
lines of highly ionized species are located compared to the optical
and near UV bands. This bump extends over the wavelength range of
$100$ to $400\,{\mbox{\AA}}$ and thus a stronger UV bump results in a line locking dynamical effect which is relatively less important. 

For comparison we also show in figure \ref{velos} the velocity profile
obtained with the Sobolev approximation for the case where the flow is
irradiated by a pure continuum emission and for the same set of initial conditions. The Sobolev approximation does not include line blanketing and, therefore, results in a larger  velocity gradient and a higher terminal velocity (by $\sim 10\%$ in this case). Our calculations show that using the Sobolev scheme to model the effect of line locking will over estimate the terminal velocity by a factor of a few since the absorption lines are exposed to an artificially enhanced flux level across the entire flow.

\begin{figure}
\includegraphics[width=84mm]{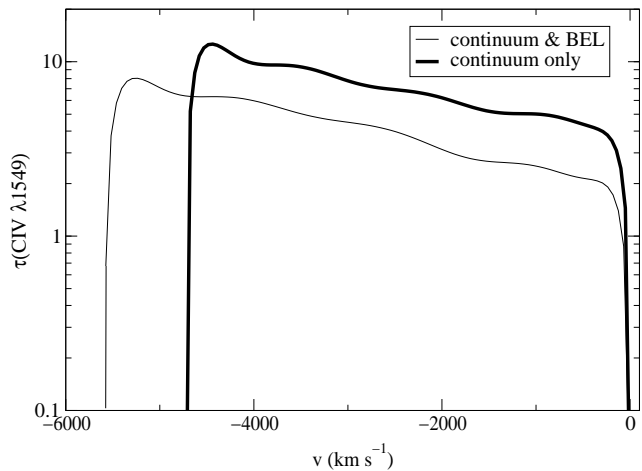} 
\caption{The velocity dependent optical depth of \civ~$\lambda 1549$  for flows accelerated by AGN continuum with and without broad emission lines. Broader profiles with  smaller optical depths are obtained when line locking is included (figure \ref{velos}). Note that the area under the curves is larger for the low velocity flow (thick curve) due to the larger column density. Note also that the optical depth is increasing with velocity  due to changes in the ionization  of carbon as a result of self-shielding.}
\label{spec_sig}
\end{figure}

The different velocity profile obtained when line locking is significant, results in a smaller flow column density since $\epsilon \propto v^{-1}$ and $N_H=\int_{r_0}^\infty  \epsilon n_H dr$. The lower column density (by roughly 50\% for the model considered here) and the larger flow velocity create a shallower absorption trough  compared to the case where line locking is discarded (figure \ref{spec_sig}).  

\begin{figure}
\includegraphics[width=84mm]{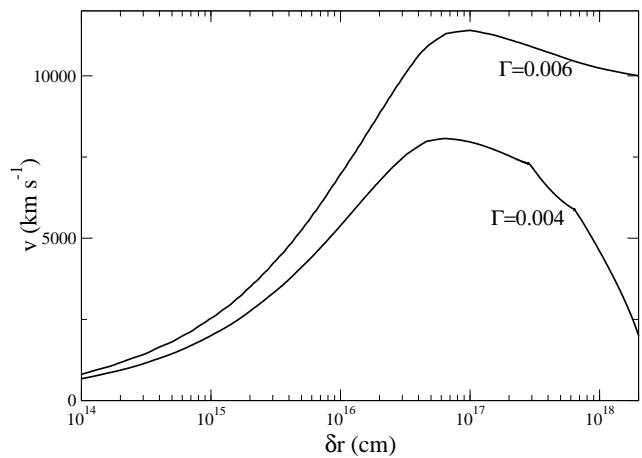} 
\caption{Velocity profiles for BAL flows with various Eddington
  ratios, $\Gamma$. The $\Gamma=0.004$ case produces a marginal
  outflow. All such profiles are very different from Sobolev type solution which are monotonically accelerating with $v(r)\propto (1-r_0/r)^{1/2}$ (see text).}
\label{gamma}
\end{figure}

The exact scheme  allows us to study a new family of continuous flow models: Non-monotonic flows. Such flows have been studied  in the framework of the cloud model (e.g., Mathews 1990 and CN01) but were excluded from wind calculations on accounts of their inconsistency with the Sobolev assumptions. Non-monotonic flows have been observed in low luminosity AGN (Ruiz et al. 2001) and are also predicted by recent hydrodynamical calculations (Proga, Stone, \& Kallman 2001). In figure \ref{gamma} we show two examples of outflows from an object shining well  below the Eddington limit. In both examples $U_x=10^{-2.5}~,r_0=10^{17}~{\rmn {cm}}$. We assume that the flows are initially rotating with a Keplerian velocity  and conserve angular momentum as they accelerate (i.e., $g(r) \propto r^{-2}(1-r_0/r)$). We study cases where $\Gamma=0.006$, and $\Gamma=0.004$  which, for the given initial conditions, imply escape velocities of $5000~{\rmn {km~s}^{-1}}$ and $7500~{\rmn {km~s}^{-1}}$ respectively. The $\Gamma=0.006$ model produces an outflow which decelerates beyond $r\simeq 2r_0$ and maintains a coasting velocity of $10,000~{\rmn {km~s}^{-1}}$.  The $\Gamma=0.004$ model corresponds to a marginal outflow which barely escapes the system ($v(r=\infty)\simeq 0$). In such flows, BAL-type velocities can be obtained for part of the time but the terminal velocity can be much lower. Thus, BAL flows can be observed in objects which shine at only a small fraction of their Eddington luminosity. 

\begin{figure}
\includegraphics[width=84mm]{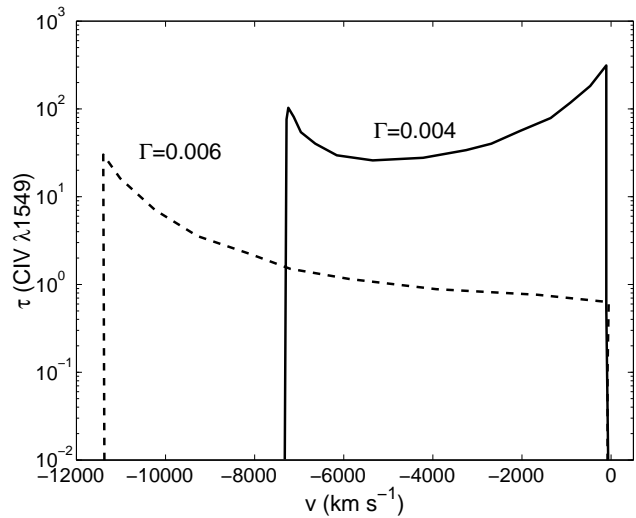} 
\caption{$\tau({\rmn {civ}}~\lambda 1549)$ as a function of velocity for the decelerating flows shown in figure \ref{gamma}. Note the large deviation from the CAK solution that predicts a constant optical depth as a function of velocity (see text).}
\label{leddp}
\end{figure}

The velocity dependence of $\tau(\mbox{\civ}~\lambda 1549)$ for the above decelerating models are shown in figure \ref{leddp}. This again is very different from the results for the monotonic BAL flows considered by Arav, Li, \& Begelman (1994; see their $\rho \propto r^{-2}$ models), and from  stellar winds profiles which are characterised by  $\tau(v)=$const.

\begin{figure}
\includegraphics[width=84mm]{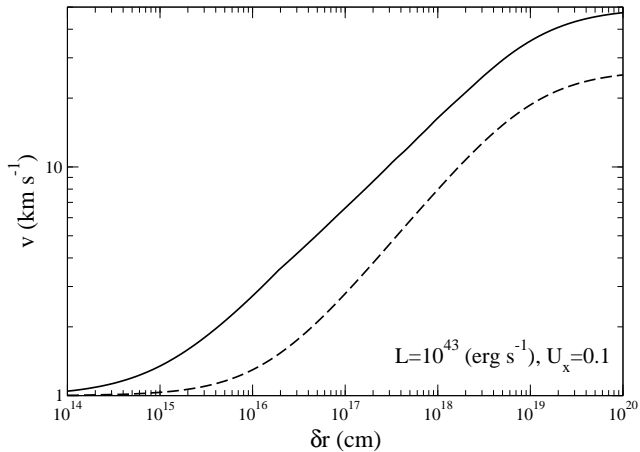} 
\caption{Velocity profiles for highly ionized NAL flow ($U_x=0.1$) for
  exact (solid line) and Sobolev (dashed-dotted line) calculations
  neglecting pressure gradient effects.  Including the skin effect can considerably enhance the terminal velocity of such flows.}
\label{nal}
\end{figure}

To illustrate the dynamical importance of the skin effect we study the
dynamics of low velocity flows. We do so by ignoring the contribution
of pressure gradients to the total force. We assume $L_{\rmn
  {tot}}=10^{43}~{\rmn {erg~s}^{-1}},~r_0=10^{19}~{\rmn {cm}}$ and
$U_x(r_0)=0.1$, typical of highly ionized gas. As shown in figure
\ref{nal}, the exact solution, that includes a larger force multiplier
close to the illuminated surface, makes a large difference to the
terminal velocity of such flows. When the flow is optically thick, the
dynamical effect can be more dramatic due to the non-linear nature of
the opacity. When initially subsonic flows are considered (see
appendix B for the case where $\rho \propto r^{-2}v^{-1}$), the skin effect will decrease the mass flow rate since larger density gradients are needed in order to balance the larger radiative force close to the illuminated surface.

The effect of $P_{\rmn {rad}}$ on the flow dynamics is small for flows
that are not very optically thick beyond the \heii~ edge since $P_{\rmn
  {rad}} \ltorder P_{\rmn {gas}}$ (EF86). The effect can be neglected once the lines become optically thin due to expansion.

\section{Conclusions and Summary}

We have presented a new method for calculating BAL and NAL flows with different properties. The main improvements over previous methods are the non-Sobolev treatment of the lines and the treatment of the internal radiation pressure.  We show that Sobolev based calculations suffer from several limitations and disagree with the exact calculations in both high and low velocity flows. The main differences are due to line blanketing in high velocity flows and incorrect calculation of the line transfer near discontinuity surfaces in low velocity flows. 

We present a new set of models pertaining to non-monotonic flows. Such
flows cannot be modeled by the Sobolev approximation, but are
important in astrophysical systems (e.g., Ruiz et al. 2001). The
calculations show that, under some geometries, it is possible to obtain high velocity flows in
objects that shine at only a few per cent of their Eddington
luminosity. Such non-monotonic flows can reach BAL velocities yet they
barely escape the system. 

We also study the effect of radiation pressure due to trapped line
photons in optically thin and optically thick differentially expanding media. In such cases, the radiation pressure distribution is considerably different from that in static uniform media. Radiation pressure is suppressed close to the flow boundaries and can obtain several local maxima. Differential expansion tends to suppress and homogenize the internal radiation pressure and reduce the dynamical effects on the flow.

This work is another step towards constructing more realistic
dynamical models for radiatively driven flows in AGN. It allows
modeling of radiative transfer in non-stationary situations  and
provides a mean for incorporating dynamics into realistic
photoionization models. The use of detailed radiative transfer codes
to model such flows is a formidable, yet highly desirable task, as it
may break the degeneracy between density and location in pure
photoionization modelling of AGN where $U_x \propto L_{\rmn {tot}} r^{-2}n_H^{-1}$. This  will allow, eventually, to constrain the mass entrained in such flows with implications to the understanding of AGN and their interaction with the host galaxy and the inter-galactic medium. 

We thank N. Arav for many enlightening  discussions. We acknowledge financial support by the Israel Science Foundation grant no. 545/00.

\appendix

\section{The Sobolev formalism}

The Sobolev approximation applies to situations where the flow properties change on scales larger than the velocity scale, e.g., $v_sdr/dv \ll \rho dr/d\rho$, where $v_s dr/dv$ is the ``Sobolev length''. The Sobolev approximation redefines the line optical depth so that each section of the flow, within one Sobolev length, is assumed to be independent on other sections. The Sobolev line optical depth (which replaces equation \ref{tau_def} in this paper) is defined as
\begin{equation}
\tau^{X_i}_{\rmn s}=\frac{0.026n^{X_i} f_l \lambda_l}{n_e
  \sigma_T v_{\rmn s}}t.
\label{tau_sobol}
\end{equation}
In the Sobolev approximation, the bound-bound force multiplier, $M^{\rmn {sob}.}_{\rmn {bb}}$, is given by,
\begin{equation}
M_{\rmn {bb}}^{\rmn {sob}.}= \sum_l \left ( \frac{
    E_l L_{E_l}e^{-\tau_{\rmn {bf}}(E_l,r)}}{L_{\rmn {tot}}} \right ) \left (
  \frac{v_{\rmn s}}{c} \right ) \frac {1-e^{-\tau^l_{\rmn s}}}{t}.
\label{mbb_sob}
\end{equation}
This definition of $M_{\rmn {bb}}^{\rmn {sob}.}$ is somewhat different from that
of previous works (e.g., Arav, Li, \& Begelman 1991, Stevens, \&
Kallman 1990; see however Proga, Stone, \& Kallman 2000)
because we include the bound-free optical depth 
term, $\tau_{\rmn {bf}}(E,r)$. This is done in order to allow a meaningful
comparison with the exact numerical method  (equations \ref{mbb} \&
\ref{tau_def}).

Ignoring the effects of self-shielding by bound-free processes, CAK provide a most convenient approximation for $M_{\rmn {bb}}^{\rmn {sob}}$,
\begin{equation}
M_{\rmn {bb}}^{\rmn {CAK}}=kt^{-\alpha},
\label{cak}
\end{equation}
where $\alpha \sim 0.9$ for AGN (Murray et al. 1995), and $k$ is a
normalization constant. This parameterization was later refined by Owocki, Castor, \& Rybicki (1988) to prevent the divergence of equation \ref{cak} at $t\ll 1$. The CAK parameterization, when plugged into the equation of motion, requires $vdv/dr=$const, which leads to the canonical stellar wind solution, $v=v_\infty (1-r_0/r)^{1/2}$.

{\section{Critical points in a non-Sobolev formalism}

Here we follow the arguments outlined in Abbott (1984) and apply them
to AGN flows. For a spherical continuous flow ($\rho \propto r^{-2} v^{-1}$), the equation of motion (equation \ref{eqnmot}) can be rearranged to give,
\begin{equation}
\frac{1}{v}\frac{dv}{dr}=\frac{1}{v^2-v_s^2} \left [ \frac{2v_s^2}{r}+\frac{\sigma_TL_{\rmn {tot}}}{4\pi r^2m_Hc} \left (M(r)-\frac{1}{\Gamma(r)} \right ) \right ],
\label{eqnmot2}
\end{equation}
where $m_H$ is the proton mass. This equation has a singular point at $v=v_s$ and a transonic solution can be obtained only if the numerator satisfies
\begin{equation}
\frac{2v_s^2}{r}+\frac{\sigma_TL_{\rmn {tot}}}{4\pi r^2m_Hc} \left ( M(r)-\frac{1}{\Gamma(r)} \right )=0
\label{nom}
\end{equation}
at the sonic point. 
Equation \ref{nom}  has one or more solutions {\it if} $M(r)<1/\Gamma(r)$. This condition is met for line driven winds and sources that shine well below their Eddington limit (e.g., stars). However, the Eddington ratio in luminous AGN is of order $0.1$ and, unless the flow is highly ionized, $M_{\rmn {bf}}+M_{\rmn {bb}}>1/\Gamma$ (e.g., CN01 and the calculations presented in this paper). Thus, a stationary (non-accelerating) hydrostatic subsonic structure cannot be formed. Furthermore, observations show that the gas temperature in AGN is {\it not} related to its location (e.g., the broad line region gas whose sound speed is $\sim 10~{\rmn {km~s}^{-1}}$ moves in Keplerian orbits with velocities in excess of $1000~{\rmn {km~s}^{-1}}$).

The above considerations suggest that, in AGN flows, the subsonic gas is never observed and therefore can be ignored. This means that the mass flow rate is determined by the initial conditions at a location where $v>v_s$. Physical scenarios which could motivate this assumption are the evaporation of cool gas to supersonic velocities from large condensations in the BLR, the central accretion disk, the torus, etc.}

\newpage
\onecolumn

{\raggedright \sloppy
  {\huge \bf  Erratum: Non-Sobolev modelling of radiation pressure
    driven flows in Active Galactic Nuclei. }
\vskip 23pt
{\LARGE Doron Chelouche and Hagai  Netzer \\}
       {\small  {\it  School of Physics and Astronomy and the Wise Observatory,
        The Beverly and Raymond Sackler Faculty of Exact Sciences,\\
        Tel Aviv University, Tel Aviv 69978, Israel}}

\author[Doron Chelouche and Hagai Netzer]
       {Doron Chelouche\thanks{email: doron@wise.tau.ac.il; netzer@wise.tau.ac.il} and Hagai
         Netzer\mbox{\raise.9ex\hbox{$\star$}} \\
        School of Physics and Astronomy and the Wise Observatory,
        The Beverly and Raymond Sackler Faculty of Exact Sciences,\\
        Tel Aviv University, Tel Aviv 69978, Israel}

\vspace{0.4in}

{The above paper (Chelouche \& Netzer 2003) includes an erroneous
reference to the work of Elitzur  
\& Ferland (1986) which resulted in an erroneous Fig. 1. Contrary to
our statement  in section 2.3, our approach to the calculation of the
escape probability in the case of a static medium is in fact similar to the
one proposed by Ferland \& Elitzur (1984) and Elitzur \& Ferland
(1986). In both methods the escape probability, $\beta_e^{\rm static}$, increases toward the
rims of the cloud (the Elitzur \& Ferland expression is $\beta_e^{\rm static}=0.5\left [ \beta_e^{\rm
    static}(\tau^+)+\beta_e^{\rm static}(\tau^-) \right ]$ where $\tau^\pm$ are the measured optical depths from points inside the cloud to
its rims). The only difference is in the exact way the radiation
pressure drops toward the edge of the cloud. Both methods are based on
the escape probability approximation and we do not consider them to be
significantly different (this is
illustrated in the corrected Fig. 1 above). We 
apologize for the error and thank G. Ferland and M. Elitzur for
pointing it out.}

\begin{figure*}
\includegraphics[width=84mm]{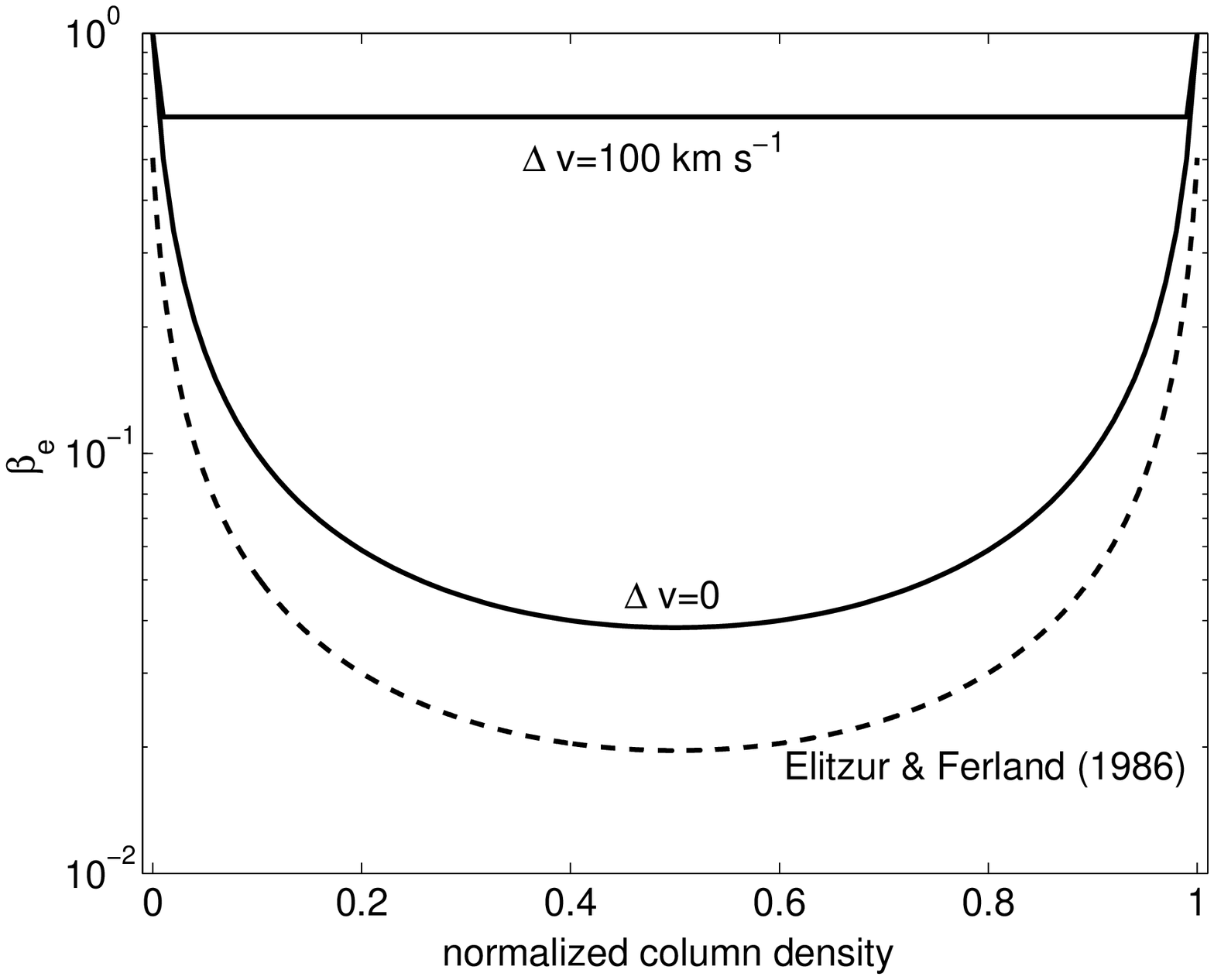} 
{\newline {\bf Figure 1:} Escape probability for a static medium showing the corrected
  Elitzur \& Ferland approximation}
\end{figure*}

\end{document}